# Enhanced Carrier Transport by Transition Metal Doping in WS$_2$ Field Effect Transistors

Maomao Liu,[1] Sichen Wei,[2] Simran Shahi,[1] Hemendra Nath Jaiswal,[1]
Paolo Paletti,[3] Sara Fathipour,[3] Maja Remskar,[4] Jun Jiao,[5]
Wansik Hwang,[6,†] Fei Yao,[2,†] and Huamin Li[1,†]

[1] Department of Electrical Engineering, University at Buffalo, the State University of New York, Buffalo, NY 14260, USA

[2] Department of Materials Design and Innovation, University at Buffalo, the State University of New York, Buffalo, NY 14260, USA

[3] Department of Electrical Engineering, University of Notre Dame, South Bend, IN 46556, USA

[4] Department of Solid State Physics, Jozef Stefan Institute, Ljubljana, 1000, Slovenia

[5] Center for Electron Microscopy and Nanofabrication, Portland State University, Portland, OR 97207, USA

[6] Department of Materials Engineering, Korea Aerospace University, Goyang 10540, Republic of Korea

----------------------------------

[†] Corresponding authors: whwang@kau.ac.kr, feiyao@buffalo.edu, huaminli@buffalo.edu

## Abstract

High contact resistance is one of the primary concerns for electronic device applications of two-dimensional (2D) layered semiconductors. Here, we explore the enhanced carrier transport through metal-semiconductor interfaces in $WS_2$ field effect transistors (FETs) by introducing a typical transition metal, Cu, with two different doping strategies: (i) a "generalized" Cu doping by using randomly distributed Cu atoms along the channel and (ii) a "localized" Cu doping by adapting an ultrathin Cu layer at the metal-semiconductor interface. Compared to the pristine $WS_2$ FETs, both the generalized Cu atomic dopant and localized Cu contact decoration can provide a Schottky-to-Ohmic contact transition owing to the reduced contact resistances by 1 – 3 orders of magnitude, and consequently elevate electron mobilities by 5 – 7 times higher. Our work demonstrates that the introduction of transition metal can be an efficient and reliable technique to enhance the carrier transport and device performance in 2D TMD FETs.

## Introduction

Tungsten disulfide ($WS_2$) with a semiconducting 2H phase is one of two-dimensional (2D) transition metal dichalcogenides (TMDs) exhibiting a series of unique properties, such as strong spin-orbit coupling, band splitting, and high nonlinear susceptibility[1-3]. Especially for future nanoelectronic applications, $WS_2$ stands out as a promising channel material compared to other 2D semiconductors. For example, $WS_2$ has a direct bandgap of 1.4 – 2.0 eV[4-7] for the monolayer and an indirect bandgap of 1.2 – 1.3 eV[4-6] for the bulk crystals. The carrier mobility of $WS_2$ has been theoretically predicated up to ~5,300 $cm^2$/Vs at 77 K[8] and ~700 – 1,100 $cm^2$/Vs at room temperature[8,9], which exceeds most of the commonly used semiconducting TMDs such as $MoS_2$

(340 cm$^2$/Vs), MoSe$_2$ (240 cm$^2$/Vs), WSe$_2$ (705 cm$^2$/Vs), owing to the relatively small effective mass (0.34$m_0$ for electrons and 0.46$m_0$ for holes, where $m_0$ is the free electron mass)[7]. Although the experimentally demonstrated electron mobilities, limited by Coulomb impurities, charge traps, surface defects and roughness, are much lower than the theoretical predication, new techniques have been developed to practically improve the mobility, for example, by exploiting $h$-BN[10] or high-$k$[11] dielectrics. For the application of field-effect transistors (FETs), monolayer WS$_2$ FETs are predicated to outperform other TMD FETs in terms of the on-state current density ($J_{D,on}$) for both p- and n-type transistors (~2,800 µA/µm for the monolayer WS$_2$ versus 2,200 – 2,400 µA/µm for the monolayer MoS$_2$, MoSe$_2$, and MoTe$_2$ FETs)[11]. In addition to the carrier mobility, the pristine hysteresis width of WS$_2$ during reliability tests is the lowest compared to MoS$_2$, MoSe$_2$ and MoTe$_2$ FETs[12]. The current on/off ratio at room temperature has been experimentally demonstrated up to ~10$^6$ for the monolayer WS$_2$ FETs[13,14] and to ~10$^8$ for the multilayer WS$_2$ FETs[15]. A nearly ideal subthreshold swing (SS) of 70 mV/decade at room temperature has been demonstrated in a simple back-gated WS$_2$ FET through a 10-nm-thick SiO$_2$ layer[16]. To further break the thermal limit of the SS, various WS$_2$-based vertical transistors operated by a band-to-band Zener tunneling mechanism have been investigated, including Gr/WS$_2$/Gr[17] and WS$_2$/SnS$_2$[18] van der Waals (vdW) heterostructures where Gr is the monolayer or multilayer graphene.

To fully explore the potential of WS$_2$ as the semiconducting channel material for the nanoelectronic device applications, the manipulation and improvement of its physical properties especially the carrier transport is highly required. For example, substitutional doping can occur as direct substitution of atoms in the lattice or into interstitial sites between existing atoms in the lattice. Owing to the unique nature of the vdW gaps, the atomic dopants in 2D materials can also intercalate between the layers, resulting in the changes in morphologic, electronic, optical, magnetic, and catalytic properties etc[19,20]. Compared to group XI elements such as Ag and Au which are widely used for wire-bonding, interconnects, and electrode materials, Cu as a representative transition metal has been demonstrated to effectively shift the Fermi level ($E_F$) up to the minima of the conduction band edge ($E_C$) and induce an n-type doping on black phosphorus (BP, or phosphorene), owing to its low electronegativity which can easily donate its 4$s$ electron to BP[21,22]. The Cu doping effect on other 2D materials, such as graphene[23], $MoS_2$[24,25], $MoSe_2$[25], $Bi_2Se_3$[26-28], $ZrSe_2$[29], $SnSe_2$[30], and SnS[31] etc. have also been investigated. Especially for $WS_2$, Cu can serve as an effective dopant to induce a microscopic ferromagnetic development which originates from the *p-d* hybridization between the Cu dopant and its neighboring S atoms and consequently the splitting of the energy levels near $E_F$[32]. However, the impact of Cu doping on the electronic carrier transport of $WS_2$ has not been explored experimentally yet. Furthermore, considering the wide applications of Cu as transistor contacts and interconnects in current semiconductor technology, the metal-semiconductor contact condition at the Cu/$WS_2$ interface critically limits the carrier injection and collection efficiency during the transistor operation, especially for the extremely-scaled short-channel devices where the contacts play a more important role in the carrier transport and device performance[33,34]. Therefore, an understanding of Cu/$WS_2$ contact can be one of the keys to achieve the full performance potential of the emerging $WS_2$ transistors.

In this work, we experimentally reveal the impact of Cu doping on the carrier transport of $WS_2$ through the metal-semiconductor interfaces, and statistically evaluate the transistor performance. Specifically, Cu as the representative transition metal is introduced using two different techniques. In the Cu-doping technique, Cu is randomly distributed into the $WS_2$ crystals during the synthesis and acts as "generalized" n-type atomic dopants within the channel. Whereas in the Cu-contact technique, Cu is deposited as an ultrathin contact decoration layer on the WS$_2$ surface and provides a "localized" doping effect only at the metal-semiconductor interface. A schematic illustration of these two doping strategies and the corresponding electron transport through the metal-semiconductor interfaces is shown in **Fig. 1**. Compared to the pristine WS$_2$ FETs, both the Cu-doped and Cu-contact WS$_2$ FETs show a clear Schottky-to-Ohmic contact improvement, due to a significant reduction of the Schottky barrier (about 28% to 60% lower) and thus the contact resistance (over 1 to 3 orders of magnitude lower). The enhanced carrier transport can be attributed to the charge transfer with Cu which changes the Fermi level (by ~0.11 eV) of WS$_2$ along the channel for the Cu-doped devices, or reduces the workfunction of Cu (by ~0.41 eV) at the metal-semiconductor interface for the Cu-contact devices. In addition, we carry out a statistical study by testing about 30 devices for each type of the WS$_2$ FETs, and demonstrate the improvement of the transistor performance including field effect mobility ($\mu_{FE}$), on/off ratio, $J_{D,on}$, and SS. Our results suggest that, owing to the doping effect, the transition metal Cu either as the generalized atomic dopant or the localized contact decoration can effectively enhance the carrier transport and device performance for 2D semiconducting WS$_2$.

## Results

**Output characteristics.** For a comparative study, we fabricate three types of typical global-back-gate transistors: pristine, Cu-doped, and Cu-contact WS$_2$ FETs. Compared to the pristine WS$_2$ FET which is the control sample in this work, the Cu-doped WS$_2$ FETs include Cu as the generalized atomic dopants during the WS$_2$ crystal synthesis. Both the pristine and Cu-doped WS$_2$ FETs have Ti/Au (10 nm/100 nm) deposited as source and drain electrodes. The Cu-contact WS$_2$ FETs have Cu/Ti/Au (2 nm/10 nm/100 nm) as the electrodes on the pristine WS$_2$ flakes where the thin Cu layer serves as a localized contact decoration at the metal-semiconductor interface. A comparison of drain current density versus drain voltage ($J_D$-$V_D$) output characteristics at various gate voltages ($V_G$) for all three types of the devices is performed at room temperature, as shown in **Fig. 2**. The pristine WS$_2$ FET possesses non-linear current-voltage (*IV*) characteristics at low $V_D$, suggesting a large Schottky barrier at the metal-semiconductor contacts. In contrast, the Cu-doped WS$_2$ FET indicates a linear *IV* relation and thus an Ohmic contact at low $V_D$. Meanwhile, current saturation is clearly obtained in the Cu-doped device, for example, at $V_D$ of 2 V for $V_G$ of 30 V. Similar to the Cu-doped WS$_2$ FET, the Cu-contact WS$_2$ FET also shows the Ohmic contact and current saturation, but the maximum $J_{D,on}$ is increased significantly up to 42 µA/µm at $V_D$ of 5 V and $V_G$ of 30 V.

To further understand the carrier transport properties, the output characteristics at the on state ($V_G$ = 30 V) are replotted in various analytical models, as shown in **Fig. S1** in Supplementary Information. In the log($J_D$) versus $V_D^{1/2}$ curves (see **Fig. S1 (a)**), a linear dependence is found at $V_D$ > 2 V for all three types of the devices, suggesting that the carrier injections subject to the Schottky emission at room temperature[35] which is described as

$$J_D \propto exp\left[\frac{-\left(\phi_b-\sqrt{\frac{q^3 V_D}{L\pi\varepsilon_0\varepsilon_r}}\right)}{k_B T}\right] \qquad (1).$$

Here $\phi_b$ is the barrier height, $q$ is the electronic charge, $\varepsilon_0$ is the permittivity of the vacuum, $\varepsilon_r$ is the relative permittivity of WS$_2$, $L$ is the channel length, $k_B$ is the Boltzmann constant, and $T$ is the temperature. Due to the identical channel material, both the pristine and Cu-contact devices possess the similar slopes (0.29 and 0.31) compared to that of the Cu-doped device (0.03). The smaller slope of the Cu-doped device also indicates a larger $\varepsilon_r$ after the generalized Cu doping in WS$_2$. **Equation 1** can be further modified as

$$J_D \propto \frac{V_D}{L} exp\left[\frac{-\left(\phi_b - \sqrt{\frac{q^3 V_D}{L\pi\varepsilon_0\varepsilon_r}}\right)}{k_B T}\right] \quad (2),$$

and a linear dependence of $\ln(J_D/V_D)$ on $V_D^{1/2}$ (see **Fig. S1 (b)**) suggests the dominance of a trap-induced Poole-Frenkel (PF) emission. To better understand the impact of the traps on the carrier transport, the data is replotted in the $\log(J_D)$ versus $\log(V_D)$ curves (see **Fig. S1 (c)**). A current saturation occurs at $V_D > 2$ V for all the devices, implying a transition of the carrier transport from trap-filled limited (TFL) mode to space-charge limited (SCL) mode[36], and the current can be described as

$$J_D \propto \frac{1}{q^{m-2}} \left(\frac{2m-1}{m}\right)^m \left[\frac{\varepsilon_0 \varepsilon_r (m-1)}{N_t m}\right]^{m-1} \frac{V_D^m}{L^{2m-1}} \quad (3),$$

where $m$ is the power factor determined from the linear slope of the $\log(J_D)$ versus $\log(V_D)$ curves, and $N_t$ is the trap density. In the TFL region ($V_D < 2$ V), the injected carriers are increased with $V_D$ but not enough to fill the traps which have an exponential distribution[37,38]. In the SCL region ($V_D > 2$ V), all the traps are filled by the injected carriers, so the subsequently injected carriers are free from the traps and fully controlled by the space charges which limit further injection of the free carriers. The Cu-contact device has a slope (0.66) which is similar with the pristine one (0.62) but different from the Cu-doped one (0.06) in the trap-free SCL region. This result indicates that the SCL current is mainly determined by the channel material rather than the contact condition. Whereas in the TFL region, both the Cu-doped and Cu-contact devices show similar slopes (1.08 and 1.02) which are different from the pristine one (2.16), suggesting the impact of the Cu doping, either along the channel or at the contact interface, on the TFL current. In the $\ln(J_D/V_D^2)$ versus $1/V_D$ curves (see **Fig. S1 (d)**), the logarithmic dependence implies that the carrier transport is dominated by direct tunneling at the low temperature or thermionic emission (TE) at the high temperature. It can be predicted that by further increasing $V_D$, a linear decay would occur which corresponds to the Fowler-Nordheim (FN) tunneling through a triangular tunneling barrier and can be described by the following equation

$$J_D \propto \frac{q^3 m_0 V_D^2}{\phi_b L^2 m^*} exp\left[\frac{-8\pi\sqrt{2m^*}\phi_b^{\frac{3}{2}}L}{3hqV_D}\right] \quad (4).$$

Here $m_0$ is the free electron mass, $m^*$ is the effective mass of electrons in WS$_2$, and $h$ is the Planck's constant. To prevent the possible damage from the current-induced joule heating, in this work we limit the sweeping range of $V_D$ only up to 5 V so the FN tunneling is excluded. The direct tunneling through a trapezoidal tunneling barrier can be further confirmed by plotting the $\ln(J_D/V_D^2)$ versus $\ln(1/V_D)$ curves (see **Fig. S1 (e)**), based on the following equation

$$J_D \propto \frac{q^2 V_D \sqrt{m_0 \phi_b}}{h^2 L} exp\left[\frac{-4\pi\sqrt{m_0 \phi_b}L}{h}\right] \quad (5).$$

The similar linear slopes of both the Cu-doped and Cu-contact devices indicate the comparable trapezoidal barrier heights at low $V_D$. Based on these analyses, we can conclude that the carrier injection at the low $V_D$ ($V_D < 2$ V) is predominated by the thermionic emission or direct tunneling for all the devices, and the current is mainly attributed to the TFL current (see **Fig. S1 (f)**). At the high $V_D$ ($V_D > 2$ V), the carrier injection is governed by the Schottky emission and the current is changed to the SCL current. The PF emission occurs at the low $V_D$ for both the pristine and Cu-doped devices, but at the high $V_D$ for the Cu-contact one.

Temperature dependence of the $J_D$-$V_D$ output characteristics is measured from 218 to 298 K for each type of the devices, and a linear relation in the Arrhenius plot, i.e., $\ln(I_D/T^2)$ versus $q/k_B T$, is obtained for various $V_G$, as shown in **Fig. 3 (a)**. Our previous work has demonstrated a gate-dependent Schottky barrier modulation for 2D TMDs[39] and the value of $\phi_b$ for a given $V_D$ is estimated from the slope of each curve, as shown in **Fig. 3 (b)**. At zero $V_G$ or the equilibrium state, the value of $\phi_b$ is obtained as 166, 59, and 0 meV for the pristine, Cu-doped, and Cu-contact WS$_2$ devices, respectively. As $V_G$ increases, a transition between the linear and exponential decay is observed, which indicates a flat-band condition across the metal-semiconductor interface. The carrier transport is dominated by the TE when $V_G < V_{FB}$, and by the direct tunneling or FN tunneling when $V_G > V_{FB}$, where $V_{FB}$ is the flat-band gate voltage. The value of $\phi_b$ at $V_{FB}$, known as the intrinsic barrier height ($\phi_{b0}$), can be used to estimate the band offsets between the metal and semiconductors. To further eliminate the effect from the applied $V_D$, $\phi_{b0}$ is plotted as a function of $V_D$, and the values at zero $V_D$ are estimated based on a linear fit for each type of the devices, as shown in **Fig. 3 (c)**. Compared to the pristine $WS_2$ FET ($\phi_{b0}$ = 123 meV), the value of $\phi_{b0}$ is reduced to 89 meV for the Cu-doped $WS_2$ FET (i.e., ~28% reduction) and to 50 meV for the Cu-contact $WS_2$ FET (i.e., ~60% reduction), suggesting a significant improvement of the metal-semiconductor contact condition. Our results are also compared with the theoretical barrier height calculated by density functional theory (DFT) for the monolayer $WS_2$ with various metal contacts[40], as shown in **Fig. 3 (d)**. The values obtained in this work are about one order of magnitude lower due to the smaller bandgap in the multilayer structure.

Based on the extracted $\phi_{b0}$ at $V_{FB}$ and $\phi_b$ at zero $V_G$, the band diagram of the metal-semiconductor interface can be estimated quantitatively, as shown in **Fig. 4**. Assuming that the work function of Ti ($\Phi_{Ti}$) is 4.33 eV, the electron affinity for the pristine and Cu-doped $WS_2$ ($\chi_{WS2,pri}$ and $\chi_{WS2,Cu}$) can be calculated through $\phi_{b0}$ at $V_{FB}$ as 4.21 and 4.24 eV, respectively. These two values are approximately identical as ideally the electron affinity is independent of the doping level. Therefore, we take their average value (4.23 eV) for the following discussion. At the equilibrium condition, the workfunction of the pristine and Cu-doped $WS_2$ ($\Phi_{WS2,pri}$ and $\Phi_{WS2,Cu}$) can be estimated as 4.4 and 4.29 eV, respectively. Therefore, the Schottky barrier lowering in the Cu-doped $WS_2$ FET can be interpreted by the electron doping which shifts the position of $E_F$ close to $E_C$ by about 0.11 eV (from 166 meV in the pristine WS$_2$ to 59 meV in the Cu-doped WS$_2$). Such n-type doping effect by Cu has also been reported on other 2D semiconducting materials[22,29]. As a comparison, the Schottky barrier lowering in the Cu-contact WS$_2$ FET is more complicated, which is mainly attributed to a synergetic interaction of exchange repulsion, covalent bonding, and charge redistribution including electron accumulation in the gap region, depletion near the interface, as well as charge density oscillations within both the metal and semiconductor[41]. Specifically, based on the average $\chi_{WS2}$ (4.23 eV) and $\phi_{b0}$ (50 meV) at $V_{FB}$, the work function of Cu ($\Phi_{Cu}$) is estimated as 4.28 eV which is lower than the known value (4.69 eV, average of Cu (100), (110), and (111)) by about 0.41 eV. This unique decrease of $\Phi_{Cu}$ is in a good agreement with the previously theoretical predication, and can be attributed to the generation of a net interfacial dipole induced by the complex charge redistribution at the interface between Cu and group-VI TMDs[41]. Meanwhile, although the WS$_2$ bulk body remains pristine, $\Phi_{WS2,pri}$ at the interface is also reduced by contacting Cu due to the charge redistribution and surface dipole generation[41]. Thus, being different from the generalized Cu doping along the channel, the localized Cu contact leads to a strong n-type doping only at the contact surface, and gives rise to the zero $\phi_b$ at zero $V_G$. Besides, previous works have experimentally and theoretically demonstrated the great potential of WS$_2$ and other 2D TMDs acting as the Cu diffusion barrier[42,43], which also evidence that the doping effect induced by the Cu contact can be confined only at the contact interface.

**Transfer characteristics.** The comparison of $J_D$-$V_G$ transfer characteristics at room temperature is shown in **Fig. 5 (a)**, from which it can be seen that the transistor performance, such as $J_{D,on}$, $\mu_{FE}$, and SS are improved by introducing Cu as either the atomic dopants or the contact decoration. For example, the maximum $J_{D,on}$ of the pristine WS$_2$ FET at $V_D$ of 1 V is 0.94 µA/µm, and this value is increased by about 5 times in the Cu-doped WS$_2$ FET (4.52 µA/µm) and by about 19 times in the Cu-contact WS$_2$ FET (17.73 µA/µm). The highest $\mu_{FE}$ for the Cu-doped WS$_2$ FET is 21.9 cm$^2$/Vs and for the Cu-contact WS$_2$ FET is 26.3 cm$^2$/Vs, which are more than 4 times larger than that of the pristine WS$_2$ device (4.9 cm$^2$/Vs). The lowest SS of the Cu-contact WS$_2$ FET is 0.5 V/decade which is about 70% reduced than that of the pristine and Cu-doped WS$_2$ devices. It is also noted that, even with the Cu atomic dopants in the semiconductor channel or the Cu contact decoration at the metal-semiconductor interface, no extra defects or traps are introduced in the devices. The effective trap density ($n_t$) can be estimated from the hysteresis of the charge neutral point ($\Delta V_{CNP}$) as $n_t = C_{ox} \cdot \Delta V_{CNP}/q$, where $C_{ox}$ is the oxide capacitance (3.84 × 10$^{-8}$ F/cm$^2$ for a 90-nm-thick SiO$_2$ layer). Because all three types of the devices have the comparable values of $\Delta V_{CNP}$ from 13 to 17 V, the calculated $n_t$ ranges from 3 × 10$^{12}$ to 4 × 10$^{12}$ cm$^{-2}$.

Based on the transfer characteristics, transconductance ($g_m$) is calculated and their temperature dependence indicates a metal-insulator transition (MIT) phenomenon, as shown in **Fig. 5 (b-d)**. The gate voltage required for inducing the MIT ($V_{G,MIT}$) in the pristine WS$_2$ FET is reduced from about 10 to 5 and 0 V by introducing Cu as the atomic dopant and contact decoration, respectively. The shift of $V_{G,MIT}$ to the negative $V_G$ indicates a Cu-induced n-type doping effect which is consistent with our previous discussion, and can be further interpreted by the frameworks of thermally activated and variable-range hopping (VRH) models[14]. For example, activation energy ($E_a$) which is corresponding to the thermal activation of charge carriers into the conduction band is extracted by fitting the sheet conductivity ($G$) with the expression $G(T) = G_0\exp(-E_a/k_BT)$, where $G_0$ is the constant. A plot of ln($G$) versus 1000/$T$ at the on state for each type of the devices is shown in **Fig. 5 (e)**. The values of $E_a$ are extracted from the linear fit of the plots, and their dependence on $V_G$ across the subthreshold and superthreshold regions (–10 to 30 V) are calculated, as shown in **Fig. 5 (f)**. A clear reduction of $E_a$ can be found in the Cu-doped and Cu-contact WS$_2$ FETs, which is consistent with the behavior of the $V_G$-dependent $\phi_b$ (see **Fig. 3 (b)**). Assuming the dependence of $E_a$ on $V_G$ equals to the dependence of $E_F$ on $V_G$, the density of the states ($DOS$) below the conduction band edge can be extracted from the expression $dE_F/dV_G = dE_a/dV_G = C_{ox}/(C_{ox}+C_t)$, where $C_t = e^2 DOS$ is the quantum capacitance. **Figure 5 (g)** shows the calculated $DOS$ as a function of $V_G$ across the insulating and metallic regimes in a comparison with the theoretically anticipated value of $DOS$ ($DOS_{\text{2D,theory}}$) which equals to $2.85 \times 10^{14}$ eV$^{-1}$cm$^{-2}$ at the effective mass $m^* = 0.34$ $m_0$[7,14]. The average $DOS$ is increased from $10^{12}$ to $10^{14}$ eV$^{-1}$cm$^{-2}$ when $V_G$ increases from the subthreshold region to the superthreshold region. Based on the VRH model[44-46], the localization length $L_{local}$ is calculated as $L_{local} = (13.8/k_B T_0 DOS)^{1/2}$, where $T_0$ is the temperature coefficient to describe the temperature dependence of $G$ as $G(T) \sim \exp(-T_0/T)^{1/3}$. **Figure 5 (h)** shows that, with the increasing $V_G$, the electron delocalization occurs and the localization length can reach up to about hundreds of nanometers. Compared to the pristine WS$_2$ FET, both the Cu doping and Cu contact can initiate the delocalization process at lower $V_G$ and potentially benefit the low-power energy-efficient device applications.

Being consistent with the MIT behavior, the electron currents (e.g., at $V_G = 30$ V or the on state) for all three types of the devices decrease with $T$ from 218 to 298 K, whereas the hole currents (e.g., at $V_G = -30$ V or the off state) increase with $T$, as shown in **Fig. 6 (a)**. The $T$-dependent electron current variation can be attributed to the dominance of lattice scattering which decreases $\mu_{FE}$ with the increasing $T$. Here $\mu_{FE}$ is defined as $(L/W)(1/C_{ox})(1/V_D)g_m$ where $L$ and $W$ are the channel length and width, respectively. It is found that $\mu_{FE}$ follows a power-law relation with $T$ as $\mu_{FE} \sim T^{-\gamma}$, where $\gamma$ is the temperature damping factor and calculated to be 0.7, 0.72, and 1.66 for the pristine, Cu-doped, and Cu-contact WS$_2$ FETs, respectively, as shown in **Fig. 6 (b)**. Our results show a good agreement with previous reports on the WS$_2$ FETs where the value of $\gamma$ ranges from 0.73 to 1.75[14]. On the other hand, the hole currents increase with $T$ due to the thermal energy assisted generation of the minority charge carriers which contribute to the leakage current at the off state.

To further confirm the enhancement of the carrier injection through the metal-semiconductor contact interface, we fabricate transmission line measurement (TLM) devices to extract the contact resistance ($R_C$) as a function of $V_G$ for each type of the devices, as shown in **Fig. 6 (c)**. At the on state, $R_C$ is calculated as $2.2 \times 10^9$, $7.4 \times 10^6$, and $1 \times 10^6$ Ωμm for the pristine, Cu-doped, and Cu-contact WS$_2$ FETs, respectively, suggesting a significant improvement (a reduction by about three orders of magnitude) of the carrier injection by introducing Cu. Even at the off state, $R_C$ also shows a reduction by one order of the magnitude with either the Cu atomic doping or Cu contact decoration. Our results are also benchmarked as a function of the channel resistivity, $\rho_{2D,channel} = (R_{total}–R_C)W/L$[34], in a comparison with the merit of the monolayer WS$_2$[14] and few-layer Cl-doped WS$_2$[47], as shown in **Fig. 6 (d)**. Here $R_{total}$ is the total resistance measured as $V_D/J_D$. It is noted that both the Cu atomic doping and Cu contact decoration can provide the lowest $R_C/R_{total}$ ratio which is required to be 20% by the International Technology Roadmap for Semiconductors (ITRS) in 2015.

## Discussion

To eliminate the possible deviation among the devices due to the difference in terms of the flake uniformity and contaminations etc., a statistical analysis based on about 30 devices of each type is carried out at room temperature, which proves the reliability and accuracy of our results.

Under the same measurement condition, the $J_D$-$V_G$ transfer characteristics of all the pristine, Cu-doped, and Cu-contact WS$_2$ FETs are compared, as shown in **Fig. S2** in Supplementary Information. The statistical analysis of the transistor performance, including $\mu_{FE}$, $J_{D,on}$, on/off ratio and SS, are summarized in **Fig. 7**. Both the median and mean values show a clear improvement by introducing the Cu atomic doping and Cu contact decoration. For example, compared to the pristine WS$_2$ FETs, the mean values of $\mu_{FE}$, $J_{D,on}$, and on/off ratio in the Cu-doped WS$_2$ FETs are about 5, 4, and 2 times increased, respectively, and the mean value of SS is over 30% reduced. The Cu-contact WS$_2$ FETs show an even better performance, including about 7 times increases in $\mu_{FE}$ and $J_{D,on}$, 6 times increases in on/off ratio, and 26% reduction in SS.

In conclusion, we carry out a comparative study among the pristine, Cu-doped, and Cu-contact WS$_2$ FETs, and demonstrate that both the Cu atomic doping and Cu contact decoration can efficiently enhance the carrier transport in WS$_2$ FETs. The charge transfer with Cu can effectively change the Fermi level of WS$_2$ along the channel for the Cu-doped devices, and considerably reduce the workfunction of Cu at the metal-semiconductor interface for the Cu-contact devices. The Schottky-to-Ohmic contact transition with the lowered contact barrier can enhance the carrier injection through the metal-semiconductor interface, and give rise to the drastic reduction of $R_C$. The statistical analysis shows a significant improvement of the carrier transport as well as the device performance in terms of $\mu_{FE}$, $J_{D,on}$, on/off ratio, and SS.

## Methods

**Material synthesis and characterization.** Both the pristine WS$_2$ and Cu-doped WS$_2$ crystals were grown by chemical vapor transport (CVT) reaction using iodine as a transport agent. A slow growth rate from the vapor phase ensures an extremely low density of structural defects in crystals, based on our previous studies[48-51]. Briefly, a silica ampoule containing WS$_2$ powder and iodine were evacuated and sealed at a pressure of 10$^{-3}$ Pa to synthesis the pristine WS$_2$. The transport reaction ran at 1060 K with a temperature gradient of 5.6 K/cm in a two-zone furnace. After three weeks of growth, the silica ampoule was slowly cooled to room temperature with a controlled cooling rate of 15 K per hour. Approximately a few percent of the starting material were transported by the reaction to form nanotube structures, and the rest of the transported material grows thin layered crystals. For the Cu-doped WS$_2$, the growth followed the same procedure but a small amount of Cu foil (0.5% by nominal weight) were added in the ampoule. Then, both the pristine and Cu-doped WS$_2$ flakes were mechanically exfoliated from the synthesized crystals and transferred onto n-type Si substrates (0.001-0.005 Ωcm) which had a 90-nm-thick SiO$_2$ layer on the top. An energy-dispersive X-ray spectroscopy (EDX) characterization was carried out to verify the existence of Cu in the synthesized WS$_2$ flakes, as shown in **Fig. S3 (a)** and **(b)** in Supplementary Information. The Cu-doped WS$_2$ showed a uniform distribution of Cu, and the weight ratio (or atomic ratio) of W/S/Cu was measured as 0.67/0.22/0.03 (or 0.30/0.57/0.04), in addition to O, C, and I in the sample. As a comparison, the pristine WS$_2$ has the weight ratio (or atomic ratio) of W/S as 0.71/0.28 (or 0.30/0.67). Both the synthesized pristine and Cu-doped WS$_2$ were also investigated by Raman spectroscopy, as shown in **Fig. S3 (c-e)** in Supplementary Information. Compared to the monolayer WS$_2$, both the peak positions and intensity ratios[52] indicated the few-layer structures of the pristine and Cu-doped WS$_2$ flakes in this work. Moreover, the $E^1_{2g}$ peaks for the pristine and Cu-doped WS$_2$ were located consistently at around 357 cm$^{-1}$, suggesting a negligible effort of Cu atomic doping on the in-plane vibrational mode of WS$_2$. In contrast, the $A^1_g$ peak was found to be softened (red-shifted) from 422.5 to 421.9 cm$^{-1}$ due to the Cu-induced electron doping. The insensitivity of the $E^1_{2g}$ peak and the pronounced red-shift of the

$A^1_g$ peak were consistent with the sign of electron doping on other TMD semiconductors such as MoS$_2$[53].

**Device fabrication and measurement.** For a comparative study, we fabricated three types of typical global-back-gate transistors: the pristine, Cu-doped, and Cu-contact WS$_2$ FETs. Both the pristine and Cu-doped WS$_2$ FETs had Ti/Au (10 nm/100 nm) deposited as source and drain electrodes. The Cu-contact WS$_2$ FETs had Cu/Ti/Au (2 nm/10 nm/100 nm) as the electrodes on the pristine WS$_2$ flakes where the thin Cu layer serves as a contact decoration at the metal-semiconductor interface. All the pristine and Cu-doped WS$_2$ flakes had the similar thickness ($t$) ranging from 4 to 8 nm, measured by atomic force microscopy (AFM). The value of $L$ was set as 1 μm for all the FETs but the values of $W$ were varying for each device. The AFM data of a selected Cu-doped WS$_2$ FET was shown in **Fig. S4** in Supplementary Information, where $t$, $L$, and $W$ were measured to be 6 nm, 1 μm, and 4 μm, respectively. The electrical characterization was performed by measuring drain current ($I_D$) at various $V_D$ and $V_G$, and the value of $I_D$ was further normalized to $J_D$ ($J_D = I_D/W$) for performance comparison. The room-temperature measurement was performed in a dark N$_2$-filled ambient environment, and the low-temperature measurement was carried out in a vacuum environment (~2 mTorr).

## Data availability

The data that support the findings of this study are available from the corresponding author on reasonable request.

## Acknowledgements

This research was partially funded by the National Science Foundation (NSF) under Grant No. ECCS-1745621 and ECCS-1944095.

## Author contributions

W.H., F.Y., and H.L. conceived and supervised the project. M.R. synthesized the 2D crystals. P.P. and S.F. prepared the 2D thin flakes. W.H. fabricated the 2D transistor devices. M.L. and P.P. fabricated the 2D TLM devices. M.L., S.S., H.N.J., and H.L. performed the electrical characterizations. M.L., S.W., F.Y., and J.J. performed the material characterizations.

## Ethics declarations

The authors declare no competing interests.

# Figures

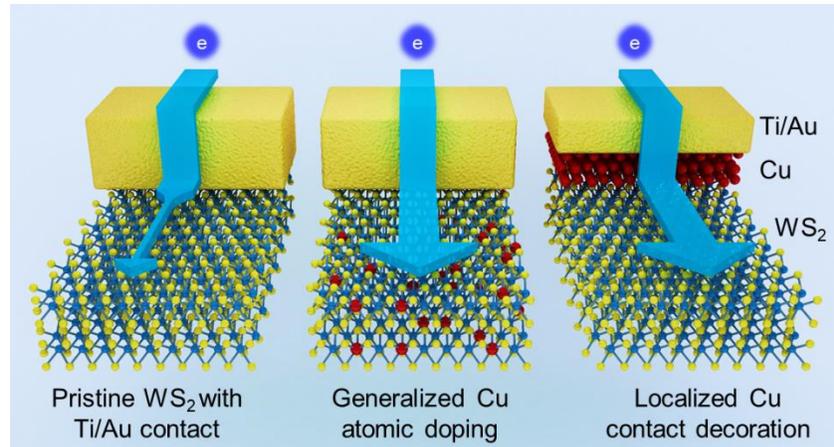

**Fig. 1** Schematic illustration of generalized Cu atomic doping and localized Cu contact decoration on $WS_2$ in a comparison with pristine $WS_2$ with Ti/Au contact. The blue arrow indicates the electron injection through the metal-semiconductor interface.

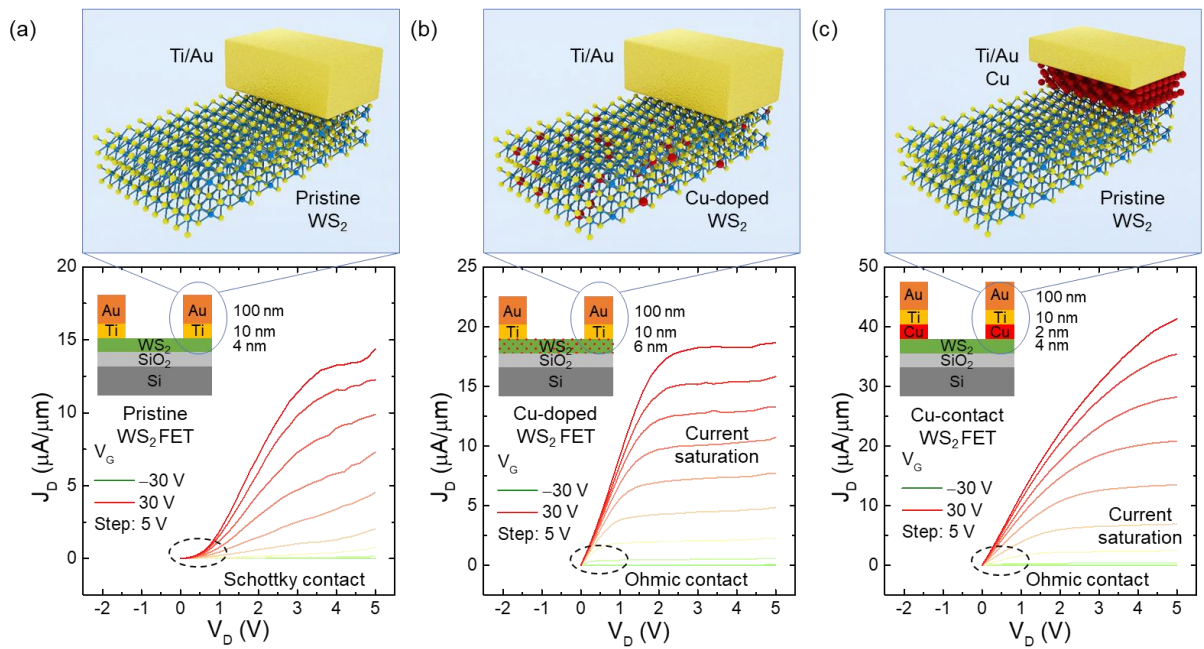

**Fig. 2** Comparison of output characteristics of WS$_2$ FETs. (a, b, c) The output characteristics of pristine, Cu-doped, and Cu-contact WS$_2$ FETs at various $V_G$ (from –30 to 30 V with a step of 5 V) at room temperature, respectively. Inset: Schematic illustration of the device structures and the corresponding metal-semiconductor interfaces.

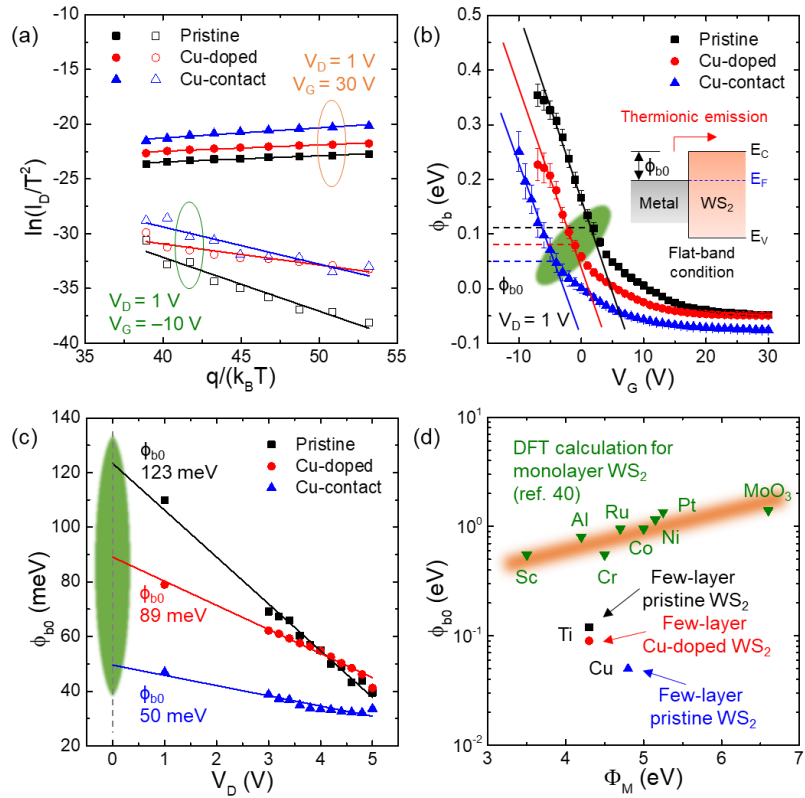

**Fig. 3** Schottky barrier heights of WS$_2$ FETs. (a) The Arrhenius plots for on state ($V_G$ = 30 V) and off state ($V_G$ = –10 V) at $V_D$ = 1 V. (b) The extracted $\phi_b$ as a function of $V_G$. The transition between linear and exponential decay highlighted in green indicates the flat-band condition at the metal-semiconductor interface. (c) The extracted $\phi_{b0}$ as a function of $V_D$. The values at zero $V_D$ are predicted by a linear fit and highlighted in green. (d) Comparison of the experimentally extracted $\phi_{b0}$ with the theoretically calculated values for various metals.

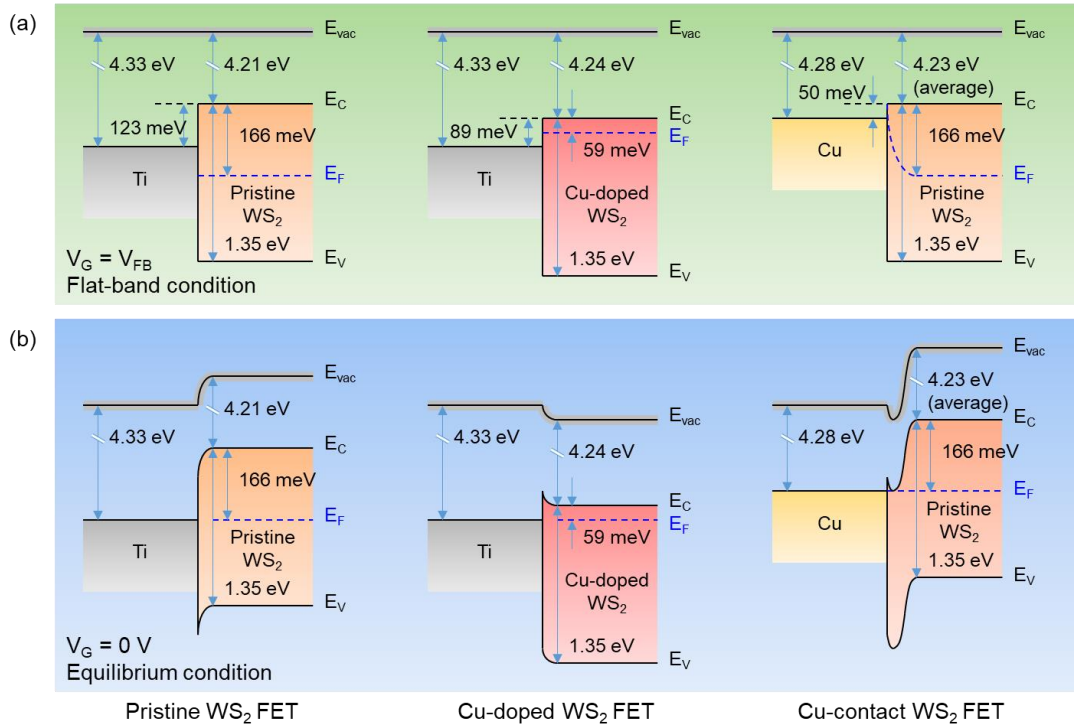

**Fig. 4** Energy band diagram of the metal-semiconductor interface at zero $V_D$. (a) Energy band diagram at flat-band condition ($V_G = V_{FB}$). (b) Energy band diagram at equilibrium condition ($V_G = 0$ V). Here $\phi_{b0}$ at $V_{FB}$ is measured to be 123, 89, and 50 meV, and $\phi_b$ at zero $V_G$ is measured to be 166, 59, and 0 meV for the pristine, Cu-doped, and Cu-contact WS$_2$ FETs, respectively. $E_{vac}$, $E_C$, $E_V$, and $E_F$ denote the vacuum energy level, the minima of the conduction band edge, the maxima of the valance band edge, and the Fermi energy level, respectively.

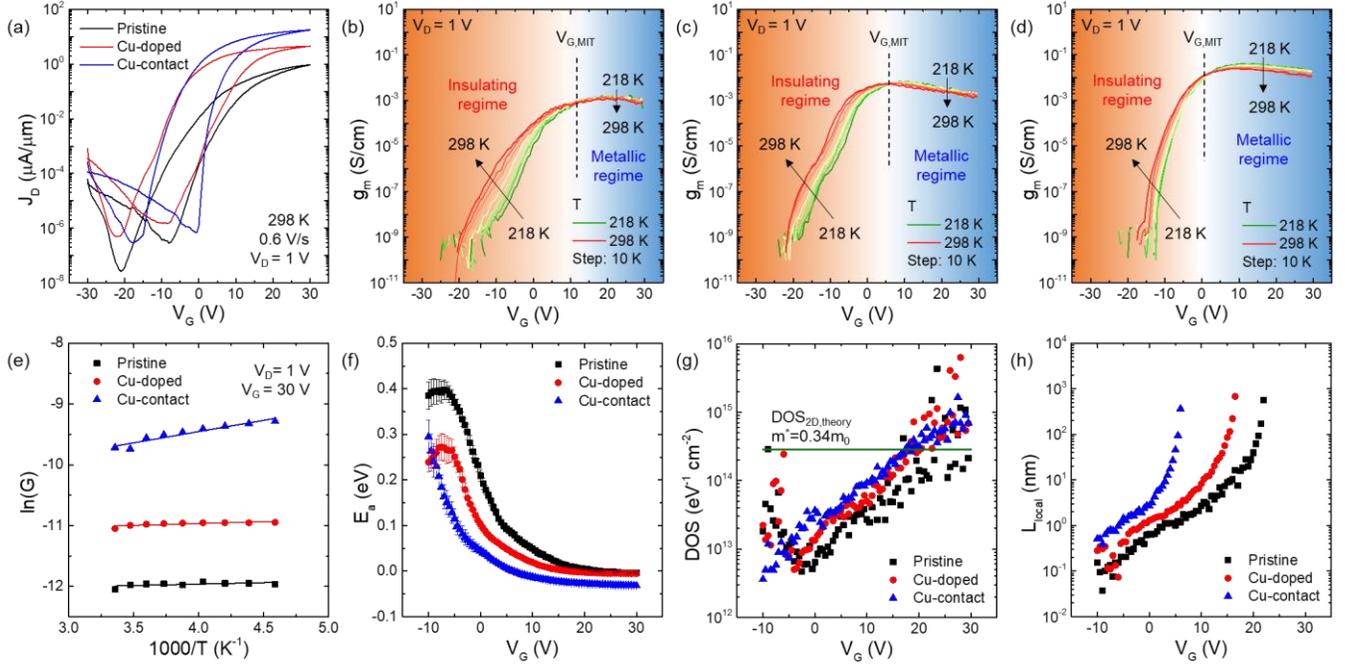

**Fig. 5** Comparison of transfer characteristics of WS$_2$ FETs and $IV$ characteristics based on the transfer curves. (a) Comparison of transfer characteristics at room temperature for the pristine, Cu-doped, and Cu-contact WS$_2$ FETs. (b-d) The extracted $g_m$ as a function of $V_G$ at various $T$ illustrates the MIT effect in the pristine, Cu-doped, and Cu-contact WS$_2$ FETs, respectively. The insulating and metallic regimes are colored in orange and blue. (e, f) A plot of ln($G$) versus 1000/$T$ shows the extraction of $E_a$ from a linear fit at $V_D$ = 1 V and $V_G$ = 30 V, and the extracted $E_a$ as a function of $V_G$ for all three types of the devices. (g, h) The extracted $DOS$ and $L_{local}$ as the functions of $V_G$ for all three types of the devices.

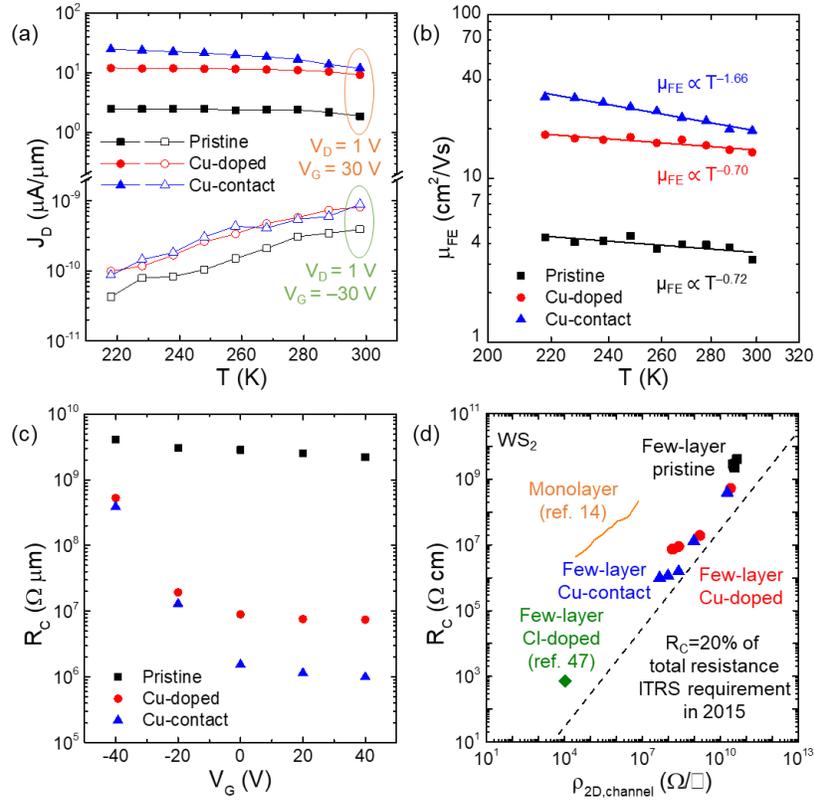

**Fig. 6** Carrier mobility and contact resistance of WS$_2$ FETs. (a, b) $J_D$ as a function of $T$ and $\mu_{FE}$ as a function of log($T$) for each type of the devices. (c, d) The extracted $R_C$ as a function of $V_G$, and its benchmarking as a function of $\rho_{2D,channel}$ in a comparison with monolayer and few-layer Cl-doped WS$_2$. The dashed line depicts the criterion $R_C = 20\%$ of total resistance in a transistor with a 22-nm-long gate as required by ITRS in 2015.

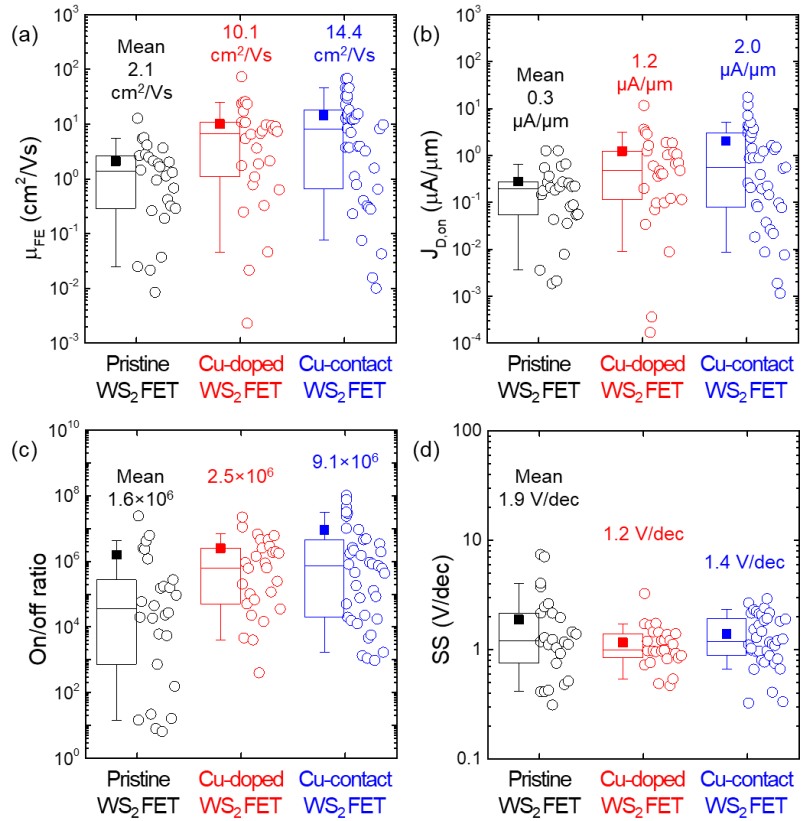

**Fig. 7** Statistical analysis of transistor performance for 26 pristine $WS_2$ FETs, 29 Cu-doped $WS_2$ FETs, and 39 Cu-contact $WS_2$ FETs. (a-d) Comparison of transistor performance including $\mu_{FE}$, $J_{D,on}$, on/off ratio, and SS, respectively. Here the box ranges from 25 to 75 percentile and the whisker ranges from 10 to 90 percentile. The bar in the box and the solid square denote the median and mean values, respectively.

# Enhanced Carrier Transport by Transition Metal Doping in WS$_2$ Field Effect Transistors

Maomao Liu,[1] Sichen Wei,[2] Simran Shahi,[1] Hemendra Nath Jaiswal,[1] Paolo Paletti,[3] Sara Fathipour,[3] Maja Remskar,[4] Jun Jiao,[5] Wansik Hwang,[6]† Fei Yao,[2]† and Huamin Li[1]†

[1] *Department of Electrical Engineering, University at Buffalo, the State University of New York, Buffalo, NY 14260, USA*

[2] *Department of Materials Design and Innovation, University at Buffalo, the State University of New York, Buffalo, NY 14260, USA*

[3] *Department of Electrical Engineering, University of Notre Dame, South Bend, IN 46556, USA*

[4] *Department of Solid State Physics, Jozef Stefan Institute, Ljubljana, 1000, Slovenia*

[5] *Center for Electron Microscopy and Nanofabrication, Portland State University, Portland, OR 97207, USA*

[6] *Department of Materials Engineering, Korea Aerospace University, Goyang 10540, Republic of Korea*

# FIGURES

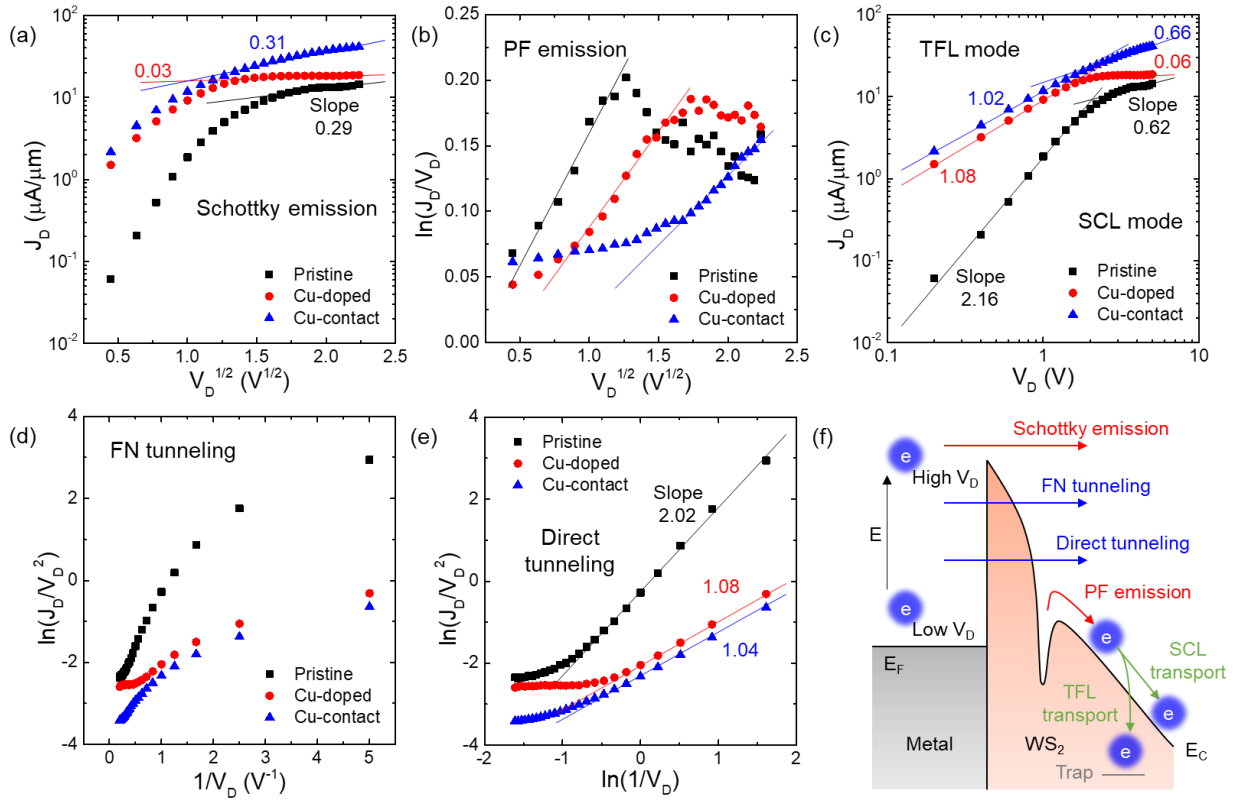

**Fig. S1** *IV* characteristics of WS$_2$ FETs at on state ($V_G$ = 30 V) based on the output curves. Various models, including (a) Schottky emission model, (b) PF emission model, (c) TFL/SCL model, (d) FN tunneling model, and (e) direct tunneling model were used for comparative investigation. (f) Energy band diagram at the metal-semiconductor interface illustrates the mechanisms of the carrier injection and transport.

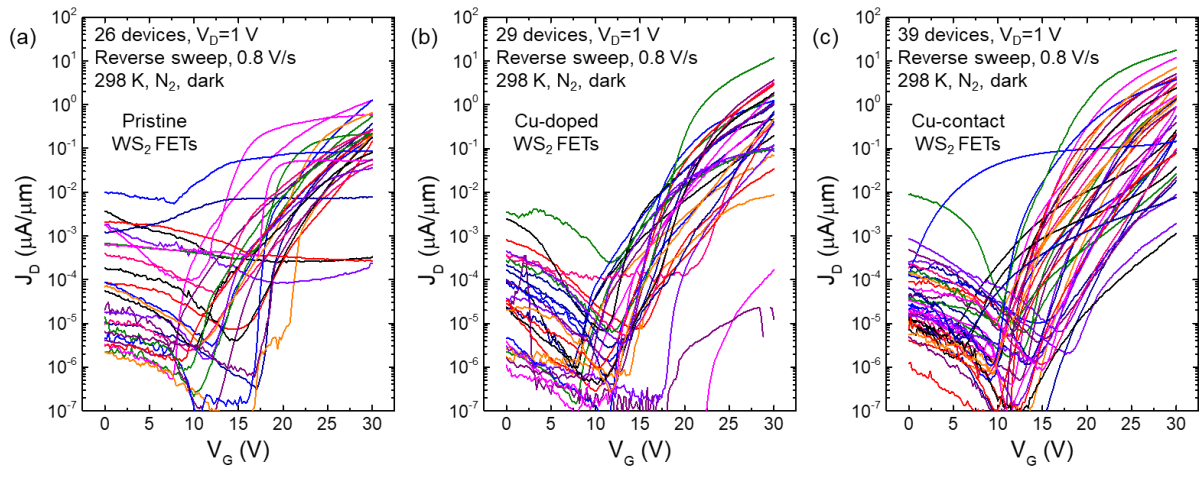

**Fig. S2** The transfer characteristics for (a) 26 pristine $WS_2$ FETs, (b) 29 Cu-doped $WS_2$ FETs, and (c) 39 Cu-contact $WS_2$ FETs.

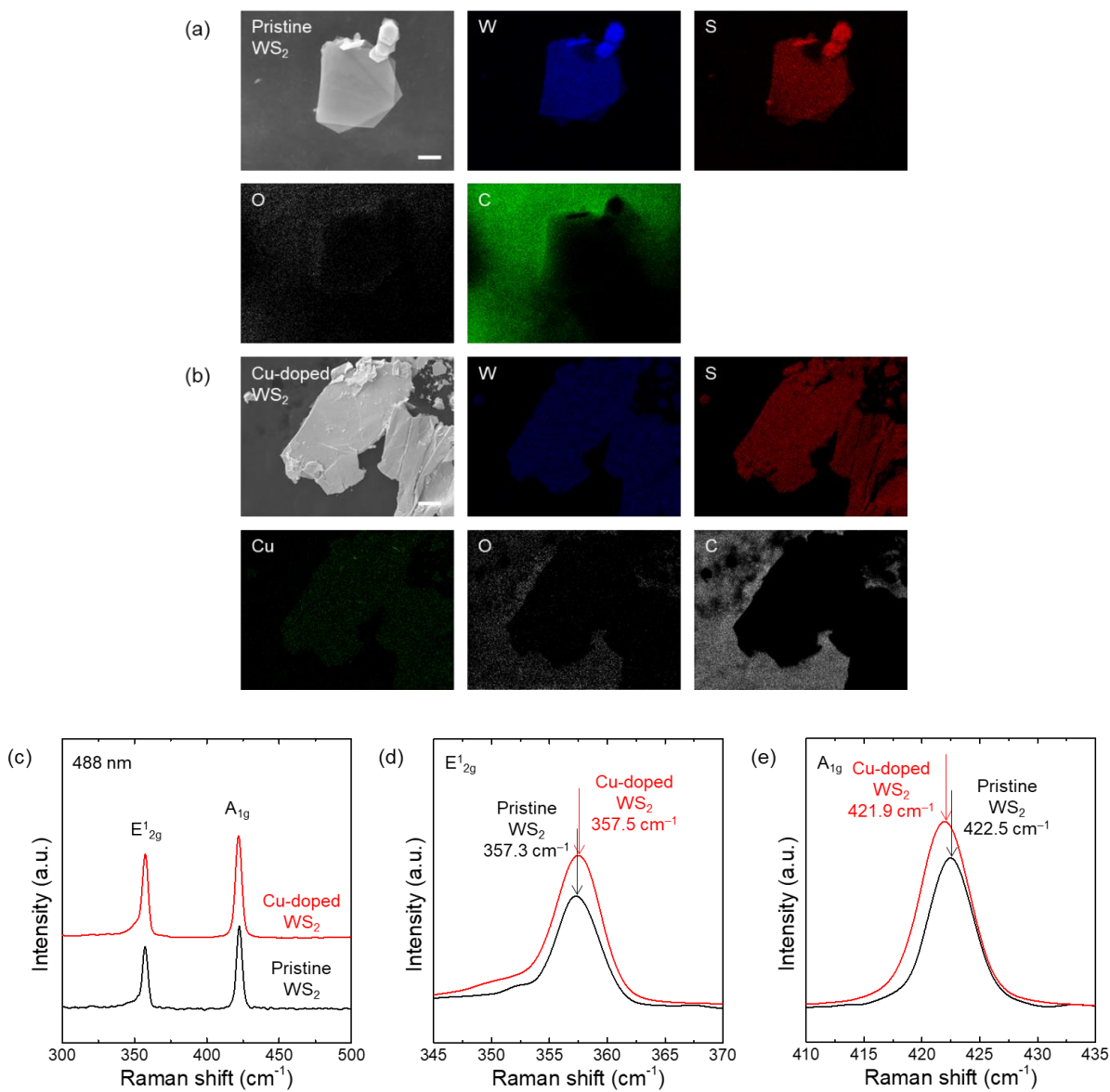

**Fig. S3** (a, b) SEM and EDX elemental mapping of a synthesized pristine WS$_2$ flake and a Cu-doped WS$_2$ flake transferred on the carbon tape surface. Scale bar: 2 μm. (c-e) Raman spectroscopy of the pristine and Cu-doped WS$_2$.

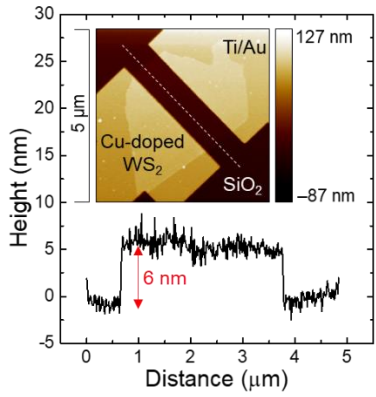

**Fig. S4** AFM mapping image of a back-gate FET using the exfoliated Cu-doped $WS_2$ flake as the channel (inset) and the cross-section profile of the channel.